\documentclass[%
 aip,
 amsmath,amssymb,
 reprint,%
]{revtex4-1}

\usepackage{graphicx}
\usepackage{dcolumn}
\usepackage{bm}
\usepackage[hidelinks]{hyperref}
\usepackage[utf8]{inputenc}
\usepackage[T1]{fontenc}
\usepackage{mathptmx}
\usepackage{etoolbox}
\usepackage{multirow}
\usepackage{amsmath,amssymb,amsfonts}
\usepackage[dvipsnames]{xcolor}

\definecolor{carminepink}{rgb}{0.92, 0.3, 0.26}

\newsavebox\myboxA
\newsavebox\myboxB
\newlength\mylenA

\newcommand*\xoverline[2][0.75]{%
    \sbox{\myboxA}{$\m@th#2$}%
    \setbox\myboxB\null
    \ht\myboxB=\ht\myboxA%
    \dp\myboxB=\dp\myboxA%
    \wd\myboxB=#1\wd\myboxA
    \sbox\myboxB{$\m@th\overline{\copy\myboxB}$}
    \setlength\mylenA{\the\wd\myboxA}
    \addtolength\mylenA{-\the\wd\myboxB}%
    \ifdim\wd\myboxB<\wd\myboxA%
       \rlap{\hskip 0.5\mylenA\usebox\myboxB}{\usebox\myboxA}%
    \else
        \hskip -0.5\mylenA\rlap{\usebox\myboxA}{\hskip 0.5\mylenA\usebox\myboxB}%
    \fi}

\makeatletter
\def\@email#1#2{%
 \endgroup
 \patchcmd{\titleblock@produce}
  {\frontmatter@RRAPformat}
  {\frontmatter@RRAPformat{\produce@RRAP{*#1\href{mailto:#2}{#2}}}\frontmatter@RRAPformat}
  {}{}
}%
\makeatother
\begin{document}
\preprint{AIP/123-QED}

\title{iGENE: A Differentiable Flux-Tube Gyrokinetic Code in TensorFlow}
\author{V. Artigues}
\affiliation{Max Planck Institute for Plasma Physics, Boltzmannstr. 2, 85748 Garching, Germany}
\affiliation{These authors contributed equally to this work}
 \email{victor.artigues@ipp.mpg.de}
\author{G. Merlo}
\affiliation{Max Planck Institute for Plasma Physics, Boltzmannstr. 2, 85748 Garching, Germany}
\affiliation{These authors contributed equally to this work}
\author{F. Jenko}
\affiliation{Max Planck Institute for Plasma Physics, Boltzmannstr. 2, 85748 Garching, Germany}%



\date{\today}

\begin{abstract}
We present iGENE, a fully-differentiable TensorFlow implementation of the electromagnetic local nonlinear gyrokinetic model, which allows us to compute gradients of any simulation output with respect to any input via automatic differentiation. We show that even if the stochastic nature of turbulence prevents the exact evaluation of gradients of nonlinear quantities of interest, they can still be successfully used to perform outer-loop tasks, such as profile predictions. This work enables the integration of gyrokinetics into automated parameter optimization, uncertainty quantification, sensitivity analysis, and AI workflows.
\end{abstract}

\maketitle

\section{Introduction}
Understanding and controlling turbulent transport in magnetized fusion plasmas is one of the central challenges on the path to commercial fusion energy. 
Micro-scale instabilities such as ion temperature gradient (ITG) modes, trapped electron modes (TEMs), and electron temperature gradient (ETG) modes drive the dominant heat and particle losses in the core of tokamaks and stellarators~\cite{jenko2000electron,jenko2002prediction,dannert2005gyrokinetic}.
Gyrokinetic simulations\cite{garbet2010gyrokinetic,horton2008drift} provide a first-principles description of these instabilities and their non-linear saturation~\cite{brizard2007foundations}, and have become an indispensable tool for interpreting experiments and guiding reactor design~\cite{hofler2025milestone,howard2021gyrokinetic,agapito2025validation,di2026first,Merlo_2026}.

Despite this success, integrating gyrokinetic simulations into predictive transport modeling workflows for optimization tasks remains computationally demanding. The standard approach couples local or global gyrokinetic calculations to 1D transport equations, iterating until the simulated fluxes balance a prescribed heat source~\cite{barnes2010direct,parker2018bringing}.
Estimating the flux sensitivity to profile gradients, i.e. the Jacobian, requires multiple independent simulations per radial location per transport step, which is expensive.
Surrogate models such as QuaLiKiz~\cite{bourdelle2016core} and neural-network accelerated variants~\cite{van2020fast} reduce this cost but sacrifice first-principles fidelity and require large training datasets.

A fundamentally different approach is to make the gyrokinetic solver itself \textit{differentiable}, so that exact gradients of any scalar output with respect to any input parameter are available at the cost of a single adjoint pass.
This capability, known as automatic differentiation (AD)~\cite{baydin2018automatic}, has recently been applied across computational physics: in fluid dynamics~\cite{bezgin2025jaxfluids}, kinetic reduced MHD~\cite{kanekar2025gandalf}, particle-in-cell simulations~\cite{ma2025jaxincell}, and 1D tokamak transport~\cite{citrin2024torax}.
We remark that having access to the gradient of any quantity of interest is much more useful than just applying it to the profile prediction task previously discussed. For example, uncertainty quantification, sensitivity analysis, control are tasks that benefit from knowledge of the gradients.
For the specific problem of profile prediction using high fidelity gyrokinetic simulations, a key open question is whether AD gradients of the \textit{non-linear} turbulent heat flux can be evaluated in a reliable and sufficiently accurate way. The turbulent system is chaotic, and the adjoint state is known to grow exponentially during backpropagation at the rate of the maximal Lyapunov exponent, eventually corrupting the gradient signal.
To the best of our knowledge, whether a useful gradient can be extracted before this divergence occurs and whether such a gradient is accurate enough to drive meaningful optimizations has not previously been investigated.

In this paper we present iGENE, a TensorFlow implementation of the electromagnetic flux-tube gyrokinetic model, and use it to address some of these questions directly.
After verifying the code in  both linear and non-linear regimes and for different kinds of microinstabilities, we address the accuracy and reliability of linear and non-linear AD gradients. As we will show, in the linear limit AD recovers the gradients exactly. Gradients of nonlinear observables are on the other hand much harder to evaluate as they indeed get unstable. Nonetheless, we find that a finite window of backpropagation steps over which the non-linear gradients remain meaningful exist and we correlate it to the turbulence autocorrelation time. As such we demonstrate that nonlinear gradients even if inaccurate are still useful and sufficient to perform practical tasks, including optimization.

The remainder of the paper is organized as follows. Section~\ref{sec:implementation} describes the gyrokinetic equations and the TensorFlow implementation. Section~\ref{sec:verification} presents linear and nonlinear verification tests. Section~\ref{sec:gradients} analyses the evaluation of the gradient and finally
Section~\ref{sec:optimization} demonstrates AD gradient-based optimization. Section~\ref{sec:conclusion} concludes. 

\section{Numerical implementation}
\label{sec:implementation}

\subsection{The gyrokinetic system}
The dynamics of the system are obtained by solving the time evolution of the gyrocenter distribution $F_s({\bf X},v_{\|},\mu,t)$ for each plasma species $s$ according to the gyrokinetic Vlasov equation:
\begin{equation}
\frac{\partial F_s}{\partial t}+\dot{\mathbf{X}}\cdot\nabla
F_s+\dot{v_{\parallel}}\frac{\partial F_s}{\partial v_\|}=0\,,
\label{eq0}
\end{equation}
with ${\bf X}$ the gyrocenter position, $\mu$ the magnetic moment and $v_{\|}$ the velocity component parallel to the background magnetic field $\mathbf{B}={\rm B}\mathbf{b}$. The equations of motion of a gyrocenter with mass $m$ and charge $q$ read
\begin{align}
\label{eq1}
\dot{\mathbf{X}}&=v_{\parallel}\mathbf{b}+\frac{B}{B^*_{\parallel}}\left(\mathbf{v}_{\nabla  B}+\mathbf{v}_{\kappa}+\mathbf{v}_{\bf E}\right), \\
\label{eq2}
\dot{v}_{\parallel}&=-\frac{\dot{\mathbf{X}}}{mv_{\parallel}}\cdot\left(\mu\nabla B +q\nabla\bar{\phi}\right) -\frac{q}{m}\dot{\bar{A}}_\parallel
\end{align}
where $\mathbf{v}_{\nabla B}=(\mu/(m\Omega B))\,\mathbf{B}\times\nabla B$ is the grad-B drift velocity, $\mathbf{v}_{\kappa}=(v_{\parallel}^2/\Omega)\,\left(\nabla\times\mathbf{b}\right)_{\perp}$
is the curvature drift velocity, and $\mathbf{v}_{\bf E}=(1/B^2)\,\mathbf{B}\times\nabla(\bar\phi-v_\|\bar{A}_\|)$ is the generalized $\mathbf{E}\times\mathbf{B}$ drift velocity. Here, $\Omega=qB/m$ is the gyrofrequency and $B^*_{\parallel}$ the parallel component of the effective magnetic field $\mathbf{B}^*=\mathbf{B}+\frac{B}{\Omega}v_{\parallel}\nabla\times \mathbf{b}+\nabla\times\left(\mathbf{b}\bar{A}_{\parallel}\right)$. Finally, $\bar\phi$ and $\bar{A}_\parallel$ are the gyroaveraged electrostatic potential and the parallel component of the vector potential. In solving Eq.~\ref{eq0} we use the delta-f method, splitting the distribution function $F$ in a fixed background $F_0$ and a fluctuating part $f$. Collisions are neglected.

The system is closed self-consistently, computing the values of $\phi$ and ${A}_{\parallel}$ by solving the Poisson equation and the parallel component of Amp\`ere's law.
The perturbed electrostatic potential is related to the perturbed charge density by means of the Poisson equation,
\begin{align}
  \nabla^2_{\perp}\phi(\mathbf{x})&=
                                    -\frac{1}{\epsilon_0}\sum_sq_s n(\mathbf{x})\nonumber\\
                                  &=-\frac{1}{\epsilon_0}\sum_s\frac{2\pi q_s}{m_s}\int{B^*_\parallel f(\mathbf{x})dv_\parallel d\mu},
\label{eq:poisson}
\end{align}
where $n(\mathbf{x})$ denotes the density perturbation of the $s$ species, expressed as the zeroth moment of the perturbed distribution function $f$ in particle space, which in the local limit reduce to multiplication with a Bessel function (see e.g. Ref.~\onlinecite{merz2008gyrokinetic} for more details). Similarly, the ${A}_{\parallel}$ potential is obtained by solving the parallel component of Amp\`ere's law for the fluctuation fields:
\begin{equation}
-\nabla^2_{\perp}A_{\|}(\mathbf{x})=\mu_0\sum_sj_{\parallel,s}(\mathbf{x}),
\label{eq:ampere}
\end{equation}
having neglected the displacement current. The perturbed parallel current $j_{\parallel,s}$ is given by the first $v_\parallel$ moment of the distribution function.
iGENE solves Eq.\eqref{eq0} in the local (flux-tube) limit~\cite{beer1995field}. Similarly to GENE, we employ a method-of-lines approach
to solve Eq.\eqref{eq0} evolving the distribution function on a fixed structured grid
in a five-dimensional phase space grid that consists of three configuration (labeled $x, y, z$) and the two velocity space $v_\parallel$ and $\mu$. Spatial coordinates are field-aligned. The underlying axisymmetry of a tokamak  corresponds to an invariance of the unperturbed system with respect to the toroidal angle, which translates to an invariance with respect to $y$. Consequently, fluctuating fields related to a linear eigenmode correspond to a single $k_y$ Fourier mode. The local limit allows the usage of periodic boundary conditions also in the radial directions, thus both $x$ and $y$ directions are implemented in Fourier space using pseudospectral
methods. The parallel direction is instead discretised with fourth-order finite differences; an equidistant grid is used in $v_\parallel$ while a Gauss-Laguerre quadrature is used in $\mu$. Finally, time integration uses a standard fourth-order Runge-Kutta (RK4) scheme. For more details we refer the reader to Ref.~\onlinecite{merz2008gyrokinetic}.

\subsection{TensorFlow implementation and automatic differentiation}

The primary innovation of iGENE is its formulation as a directed acyclic graph (DAG) in TensorFlow~\cite{tensorflow2015-whitepaper}. 
Each RK4 step corresponds to a node in the graph, so the entire simulation history is available to the framework's reverse-mode AD engine.
Given a scalar objective $J$, such as the instantaneous heat flux $Q$, the gradient with respect to any input parameter $\theta$ is computed as
\begin{equation}
\nabla_\theta J = \frac{\partial J}{\partial \mathbf{u}_T}
                  \frac{\partial \mathbf{u}_T}{\partial \mathbf{u}_{T-1}}
                  \cdots
                  \frac{\partial \mathbf{u}_1}{\partial \theta},
\end{equation}
where $\mathbf{u}_t$ is the plasma state at time step $t$. This is mathematically equivalent to the continuous adjoint method but is constructed and evaluated automatically by the framework, with no manual derivation required.

Backpropagating through all $T$ steps is memory-prohibitive in practice. Instead, we unroll only the last $N$ steps from a starting state $F_s^{t_0}$ in the saturated regime and compute the approximate gradient
\begin{equation}
\nabla_\theta^N Q = \nabla_\theta\,\hat{Q}\!\left(
  \underbrace{\xi \circ \cdots \circ \xi}_{N}(F_s^{t_0})\right),
\end{equation}
where $\xi$ denotes one RK4 step and $\hat{Q}$ extracts the heat flux.
Whether $\nabla_\theta^N Q$ converges to the true gradient of the time-averaged $Q$ as $N$ increases, and over what range of $N$ it is reliable, is the central empirical question of Section~\ref{sec:gradients}.

The same code runs on CPUs, GPUs and APUs without modification. For large simulations, iGENE uses MPI to parallelize over plasma species and the $\mu$ grid, distributing the 5D distribution function across multiple GPU nodes.

\section{Code verification}
\label{sec:verification}

Before studying its gradient properties, we establish that iGENE faithfully
reproduces the physics of GENE across the full range of regimes relevant to
this work.

We benchmark iGENE against the standard Fortran GENE version \cite{jenko2000electron,Germaschewski} for the linear growth rate $\gamma$ and real frequency $\omega$ of ITG and ETG modes using standard CBC-like parameters~\cite{dimits2000comparisons} in $s$-$\alpha$ geometry. The ITG and ETG cases share the magnetic geometry $q = 1.4$, $\hat{s} = 0.796$, $\epsilon = 0.18$, and the normalized density gradient $\omega_n = 2.2$. Here $q$ is the safety factor, $\hat{s}$ the magnetic shear and $\epsilon$ the local inverse aspect ratio. The parallel and velocity grids use $n_z = 32$, $n_{v_\parallel} = 32$ and $n_\mu = 8$ points with $L_v = 3.0$ and $L_\mu = 9.0$. The radial direction uses $n_x = 15$ modes.

For the ITG case we use $\omega_T = 6.96$ and assume adiabatic electrons. The ETG benchmark considers instead only electrons with a mass ratio $m_e/m_i = 2.7244\times10^{-4}$ (ions are adiabatic) with the same $\omega_T$, and a small but finite $\beta = 10^{-4}$.
Finally, a third benchmark case considers the onset of kinetic ballooning modes (KBM)~\cite{wilms2025implementation} by varying $\beta$ with two fully kinetic species ($\omega_{T_i} = \omega_{T_e} = 6.96$, $\omega_n= 2.22$) on a finer grid of $32\times32\times64\times16$ points in z, $v_{\parallel}, \mu$.

\begin{figure}
    \centering
    \includegraphics[width=0.90\linewidth]{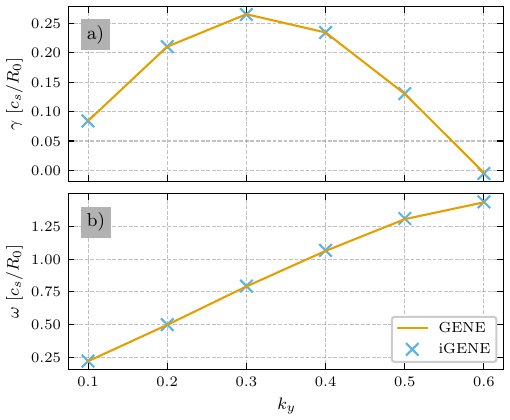}
    \caption{Linear (a) growth rate $\gamma$ and (b) real frequency $\omega$ vs.\ poloidal wavenumber $k_y\rho_s$ for the ITG benchmark.}
    \label{figure:linear_ITGscan}
\end{figure}

\begin{figure}
    \centering
    \includegraphics[width=0.90\linewidth]{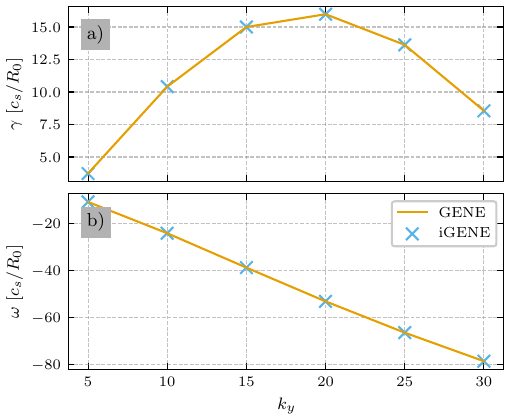}
    \caption{Linear (a) growth rate $\gamma$ and (b) real frequency $\omega$ vs.\ poloidal wavenumber $k_y\rho_s$ for the ETG benchmark.}
    \label{figure:linear_ETGscan}
\end{figure}

\begin{figure}
    \centering
    \includegraphics[width=0.90\linewidth]{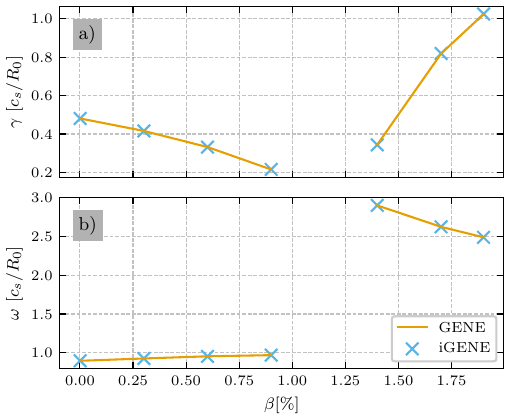}
    \caption{Linear (a) growth rate $\gamma$ and (b) real frequency $\omega$ vs.\ plasma $\beta$. The KBM branch is visible at large $\beta$
    for both codes.}
    \label{figure:linear_KBMscan}
\end{figure}

As shown in Figs.~\ref{figure:linear_ITGscan}--\ref{figure:linear_KBMscan}, iGENE reproduces the GENE growth rates and real frequencies across all three scans without discernible error. This confirms that the linear operators, field solver, and electromagnetic terms are all correctly implemented in the computational graph.

\begin{figure}
    \centering
    \includegraphics[width=0.90\linewidth]{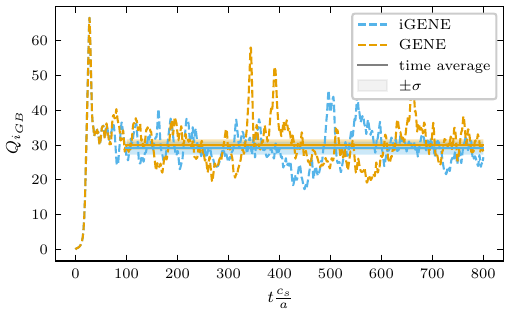}
    \caption{Ion heat flux $Q_{i,\mathrm{GB}}$ vs.\ time for the CBC with adiabatic electrons, comparing iGENE and GENE. Horizontal lines and shaded areas indicate time averages and standard deviations.}
    \label{figure:flux_CBC_trace}
\end{figure}

\begin{figure}
    \centering
    \includegraphics[width=0.90\linewidth]{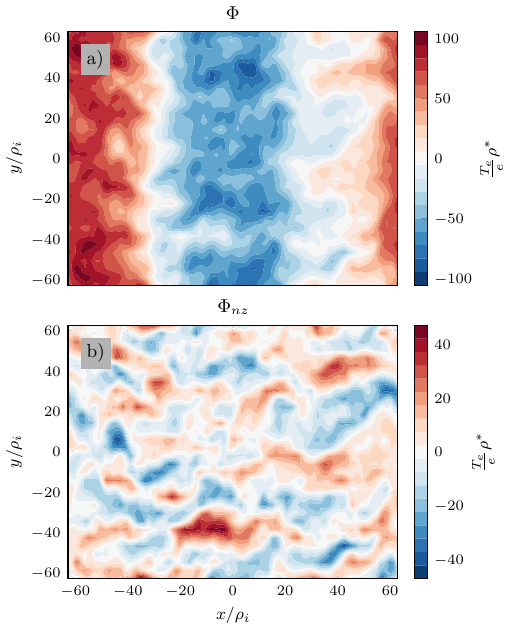}
    \caption{Contour plots of the electrostatic potential $\Phi$ (top) and the non-zonal component $\Phi_\mathrm{nz}$ (bottom) in the perpendicular $(x,y)$ plane at a representative time in the saturated phase (simulation results from iGENE).}
    \label{figure:CBC_contour}
\end{figure}

Having verified the linear physics, we turn to the non-linear saturated state, which is where the gradients studied in Section~\ref{sec:gradients} are computed. Figure~\ref{figure:flux_CBC_trace} shows the time trace of the ion heat flux $Q_i$ in gyro-Bohm units for the standard CBC with adiabatic electrons for both codes. The time-averaged values are $29.13\pm1.88$ (iGENE) and $30.17\pm1.46$ (GENE), where the uncertainty is the standard deviation over five consecutive averaging windows splitting the steady-state region evenly.
The turbulent fluctuation level is well reproduced. Figure~\ref{figure:CBC_contour} shows snapshots of the electrostatic potential $\Phi$ and its non-zonal component $\Phi_\mathrm{nz}$ in the perpendicular plane at the outboard midplane; the characteristic streamer and zonal-flow structures are correctly captured.

\begin{figure}
    \centering
    \includegraphics[width=0.90\linewidth]{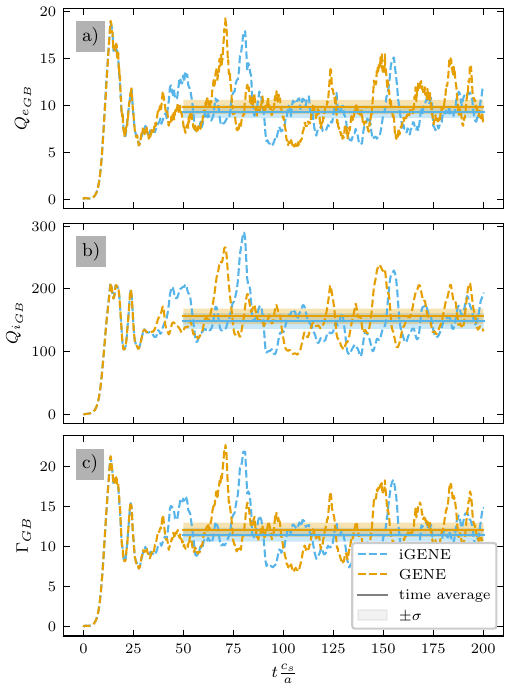}
    \caption{Heat and particle flux time traces for the CBC with kinetic electrons. (a) Electron heat flux $Q_{e,\mathrm{GB}}$, (b) ion heat flux $Q_{i,\mathrm{GB}}$, (c) particle flux $\Gamma_\mathrm{GB}$ in gyro-Bohm units, comparing iGENE and GENE.}
    \label{figure:flux_CBCE_andparticle_trace}
\end{figure}

Adding kinetic electrons introduces a second transport channel as well as a finite particle flux. Figure~\ref{figure:flux_CBCE_andparticle_trace} shows the electron heat, ion heat flux, and particle fluxes obtained from both codes.
In this case, the time-averaged electron (resp. ion) heat flux is $9.32\pm0.68$ (resp.$148.8\pm12.52$) for iGENE and $9.84\pm0.76$ (resp. $156.3\pm12.34$) for GENE; the particle flux is $11.52\pm0.87$ (iGENE) vs.\ $12.14\pm0.88$ (GENE).
All values agree within one standard deviation, demonstrating that the kinetic electron physics are faithfully reproduced as well. With the fidelity of iGENE established across linear and non-linear regimes, we now turn to its new capabilities.

\section{Evaluation of Gradients}
\label{sec:gradients}

While we verify the correctness of the gradients both for linear and non-linear simulations,  the central question of this paper is whether automatic differentiation can provide useful gradients of a quantity of interest in linear and non-linear gyrokinetic simulation.
The difficulty is fundamental: the turbulent system is chaotic, and the adjoint state that carries gradient information backwards in time is subject to the same Lyapunov instability as the forward state~\cite{lea2000sensitivity}.  In practice this means that, for non-linear simulations, $\nabla_\theta^N Q$ may converge to the true gradient for moderate $N$ but will eventually diverge as $N$ increases. Whether the convergent window is wide enough to contain a useful gradient, and which parameters are well-resolved within it, must be determined empirically.

\subsection{Linear simulations}
Once more, we begin assessing the accuracy of the AD-computed gradients for five parameters, $\theta \in \{\omega_T, \omega_n, q, \hat{s}, \epsilon\}$, starting from the linear regime. We compare AD to finite-difference (FD) reference gradients, which are obtained from a spline fit over the corresponding parameter scan. 
 
\begin{figure}
    \centering
    \includegraphics[width=0.90\linewidth]{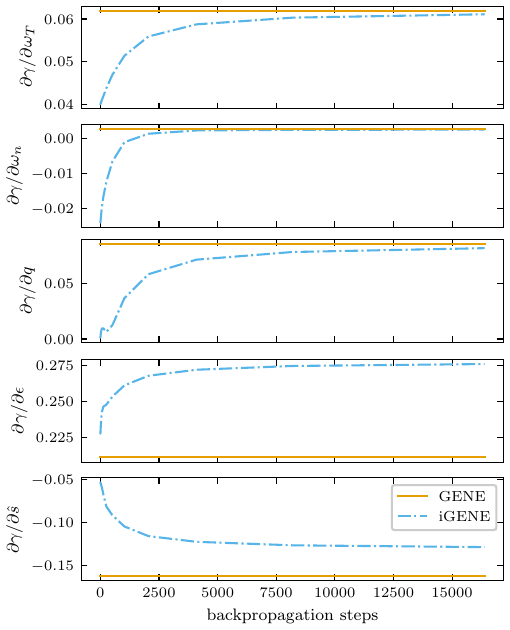}
    \caption{AD gradients as a function of backpropagation steps $N$ for a linear ITG simulation in $s$-$\alpha$ geometry, compared to FD reference values (solid lines).}
    \label{figure:linear_gradients}
\end{figure}

Figure~\ref{figure:linear_gradients} shows $\nabla_\theta^N \gamma$ as a function of $N$ for the ITG scenario and $s$-$\alpha$ geometry.
The gradients with respect to $\omega_T$, $\omega_n$, and $q$ converge monotonically to their FD values, confirming that the truncated-backpropagation approximation is valid for these parameters.
This also validates the overall AD infrastructure: in a linear, non-chaotic system there is no reason for the gradient to diverge, and none is observed.

\begin{figure}
    \centering
    \includegraphics[width=0.90\linewidth]{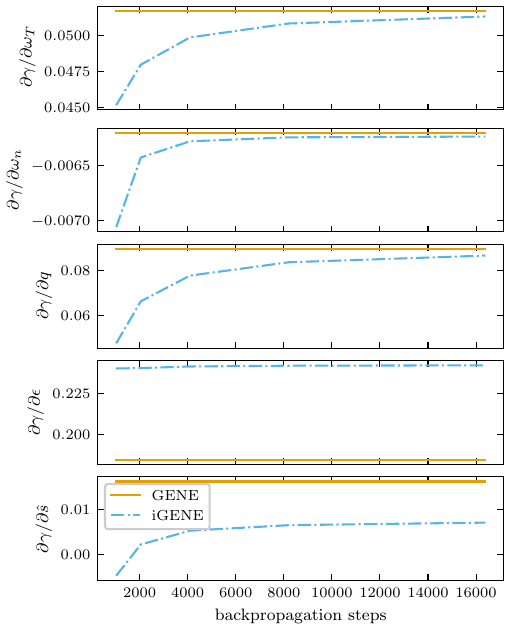}
    \caption{AD-computed gradients as a function of backpropagation steps $N$ for a linear simulation in circular geometry, compared to FD reference values (solid  lines).}
    \label{figure:linear_gradients_circular}
\end{figure}

We observed that two parameters seem to not converge to their FD values. The gradient for $\epsilon$ saturates approximately $30\%$ below the reference.
Repeating the test in circular geometry, which removes the $\epsilon \to 0$ approximation inherent to $s$-$\alpha$~\cite{lapillonne2009clarifications}, yields the same discrepancy (Fig.~\ref{figure:linear_gradients_circular}), ruling out a geometry modeling error as the cause.
The origin of the offset remains an open question. We speculate this as a consequence of the limited number of steps over which we can perform backward propagation. With the current hardware architecture, we can at most unroll 16'000 steps on one GPU due to the enormous memory requirements. A practical solution would be using forward differentiation, which however is not available in the TensorFlow. Alternative autodifferentiable languages such as JAX provide this capability; converting our solver is left for future work.  

\begin{figure}
    \centering
    \includegraphics[width=0.90\linewidth]{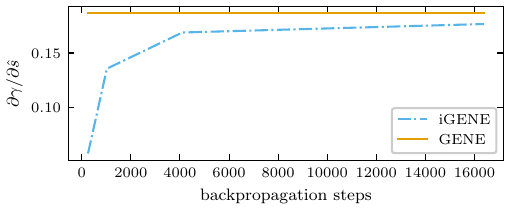}
    \caption{Linear gradient with respect to the shear $\hat{s}$ in circular geometry at $\hat{s}=0.5$.}
    \label{figure:linear_gradients_circular_shat05}
\end{figure}

\begin{figure}
    \centering
    \includegraphics[width=0.90\linewidth]{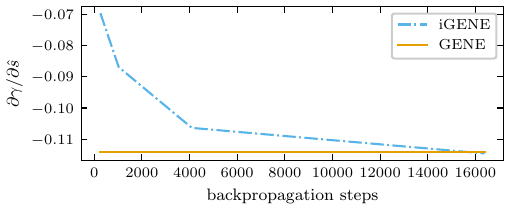}
    \caption{Linear gradient with respect to the shear $\hat{s}$ in circular geometry at $\hat{s}=1.0$.}
    \label{figure:linear_gradients_circular_shat10}
\end{figure}

The AD gradient for $\hat{s}$ shows a persistent discrepancy of approximately $30$\% relative error in $s$-$\alpha$ geometry and even $50$\% in circular geometry, relative to FD. To understand its behavior we evaluate it at two additional points $\hat{s}=0.5$ and $\hat{s}=1.0$, see figure~\ref{figure:linear_gradients_circular_shat05} and~\ref{figure:linear_gradients_circular_shat10}.
While some error remains, the relative error reduces significantly, to less than $10\%$ suggesting that the AD gradient for $\hat{s}$ retains the correct sign and approximate scaling even if the magnitude is not exact. This is understood as a consequence of not having included in our differentiable framework the mesh itself, and in particular the fact that the magnetic shear sets the radial modes $k_x$ for a given $k_y$ as a consequence of the twist-and-shift parallel boundary condition \cite{beer1995field}.

\begin{figure}
    \centering
    \includegraphics[width=0.90\linewidth]{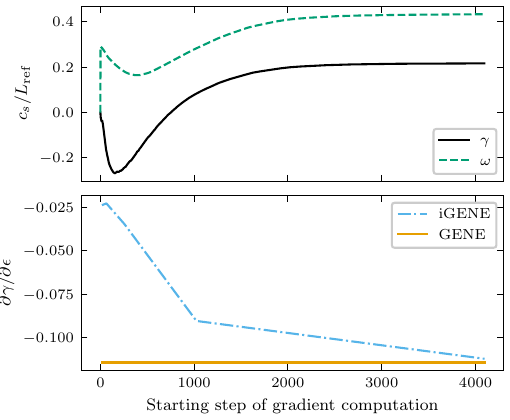}
    \caption{AD-computed gradient of the growth-rate with respect to $\epsilon$ as a function of the backpropagation starting-point. The solid line shows the FD reference value. Gradients computed from the initial transient are unreliable; they stabilize once the simulation is in the saturated phase.}
    \label{figure:linear_gradients_circular_trpepsstart}
\end{figure}

One practical question is whether backpropagation should or could begin from any point in time. To address this question, we computed the gradient for a fixed window of $N = 8000$ steps, sliding the starting point from the initial transient into the saturated state.
As shown in Fig.~\ref{figure:linear_gradients_circular_trpepsstart}, gradients computed from an unconverged state are entirely unreliable; the gradient only stabilizes once the simulation has converged. 
This establishes a key operational requirement: the simulation must reach a turbulently saturated state before backpropagation begins.

\subsection{Non-linear simulations}
\begin{figure}
    \centering
    \includegraphics[width=0.90\linewidth]{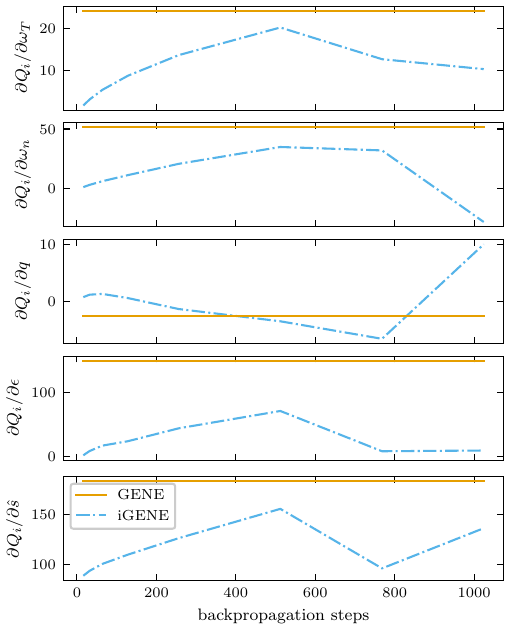}
    \caption{AD gradients of the non-linear ion heat flux $Q_i$ with respect to five parameters as a function of the number of backpropagation steps $N$. Solid lines show FD reference values. All gradients initially approach the reference before diverging at $N \gtrsim 512$, consistent with Lyapunov instability.}
    \label{figure:CBC_gradients}
\end{figure}

We now turn to the more challenging evaluation of non-linear gradients. We remark that in this case the comparison with FD is inherently more complicated due to statistical uncertainty.
Fig.~\ref{figure:CBC_gradients} shows the evolution of $\nabla_\theta^N Q_i$ as a function of $N$ for all five parameters in the CBC case. 
The behavior is qualitatively different from the linear case: each gradient initially moves toward the FD reference value but then diverges for $N \gtrsim 512$.
This is the signature of Lyapunov instability. The adjoint state amplifies exponentially at the rate of the maximal Lyapunov exponent, eventually destroying the gradient signal~\cite{lea2000sensitivity}.

The autocorrelation time for the heat flux is between $500$ and $1000$ steps depending how it is defined (as the ACF crossing $1/e$ or $0$).
Being consistent with the onset of the divergence, this suggests that the usable gradient window is set by the dynamical memory of the turbulence, though we have not proven this connection rigorously.

Within the convergent window, at $N \approx 512$, the gradients for $\omega_T$, $\omega_n$, and $q$ reach between $15\%$ and $34\%$ of their FD values, while $\epsilon$ reaches approximately $50\%$. These are not exact, but the key question for applications is not accuracy per se but whether the gradients point in a useful descent direction. We address this in the following section through various optimization tests.

\section{Gradient-based optimization}
\label{sec:optimization}

The gradient analysis of the previous section established that non-linear AD gradients are imperfect but directionally meaningful within a finite backpropagation window. We now test whether this is sufficient for  optimization problems of increasing complexity, using the Adam optimizer~\cite{kingma2014adam} throughout. In each case, gradients are computed with $N$ chosen much smaller than the convergence window identified above, and a post-optimization validation run at the final parameters confirms the result independently.

\begin{figure}
    \centering
    \includegraphics[width=0.90\linewidth]{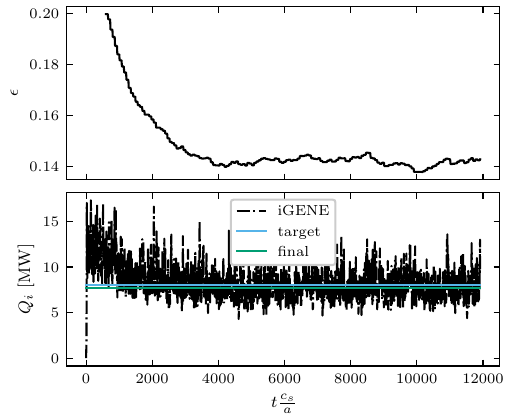}
    \caption{Single-point optimization of $\epsilon$ to achieve a target ion heat flux $Q_i = 8$ (solid line). Top: evolution of $\epsilon$ during gradient descent. Bottom: heat flux time trace (dashed-dotted line) during the final simulation at the converged value of $\epsilon$, with the post-optimization validation run confirming the result.}
    \label{figure:flux_flow_trpeps_trace}
\end{figure}

As a first test, we vary $\epsilon$ (picked for having the worst accuracy both in linear and non-linear tests) at a single radial location ($r/a = 0.6$, with parameters from the Case~2 of Parker et al.~\cite{parker2018bringing}) to match a target ion heat flux of $Q_i = 8$. Despite the large error in $\nabla_\epsilon^N Q_i$ identified in Section~\ref{sec:gradients}, the optimizer converges reliably to the target, and the post-optimization validation run confirms the result (Fig.~\ref{figure:flux_flow_trpeps_trace}). This demonstrates that a directionally correct gradient, even with imperfect magnitude, is sufficient for scalar flux matching.

We next optimize the entire temperature gradient $\omega_T$ simultaneously at seven radial positions $r/a \in \{0.2, 0.3, 0.4, 0.5, 0.6, 0.7, 0.8\}$, with a fixed temperature boundary condition at $r/a = 0.8$, targeting the prescribed heat source profile from Case~2 of Parker et al.~\cite{parker2018bringing}. The temperature profile is obtained by integrating the optimized gradients inward from the boundary. This is a substantially harder problem than the single-point case: seven optimizers must navigate the parameter space with noisy, imperfect gradients, while each simulation is running continuously and communicating temperature-gradient information to the others at regular intervals. This approach bypasses iteratively solving the transport equation~\cite{barnes2010direct,parker2018bringing}. 
Instead, the gradient of the objective (residuals of the flux and target flux, for each radial points) with respect to the temperature gradients is computed by the AD engine in a backward pass.

Each of the seven flux-tube simulations uses a grid of $128\times24\times24\times32\times8$ points in $(k_x, k_y, z, v_\parallel, \mu)$, with $L_v = 3.0\,v_{t,s}$, $L_\mu = 9.0\,T_s/B_\mathrm{ref}$, $L_x \approx 119\,\rho_s$, $k_{y,\mathrm{min}} = 0.05$, $R = 3.0$\,m, $a = 1.0$\,m, $\eta_z = 2.0$, $\eta_v = 0.2$, and $\Delta t = 8\times10^{-3}\,R/c_s$.
Optimization begins after $6\times10^4$ steps of initial saturation, using Adam with $\beta_1 = \beta_2 = 0.222$ and learning rate $0.1$.
Gradients are unrolled over $N=16$ steps every $1000$ simulation steps, clipped at threshold $0.2$, and averaged over $6$ evaluations before each parameter update.
The time step is reduced by a factor of $0.8$ during the backward pass.
The full run covers $1.6\times10^6$ steps across 7 nodes (2 GPUs per radial position), chosen to verify long-term stability rather than as a minimum requirement.

\begin{figure}
    \centering
    \includegraphics[width=0.90\linewidth]{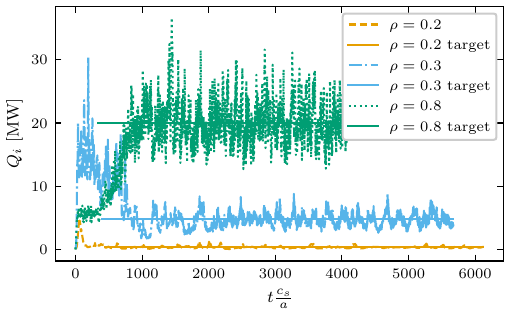}
    \caption{Time traces of the ion heat flux $Q_i$ at three radial positions ($r/a = 0.3$, $0.5$, $0.8$) during the optimization for a simulation with adiabatic electrons. Optimization begins at $t \approx 380$ (after $6\times10^4$ steps); target fluxes are shown as horizontal lines of matching color.}
    \label{figure:flux_flow_2_trace}
\end{figure}

\begin{figure}
    \centering
    \includegraphics[width=0.90\linewidth]{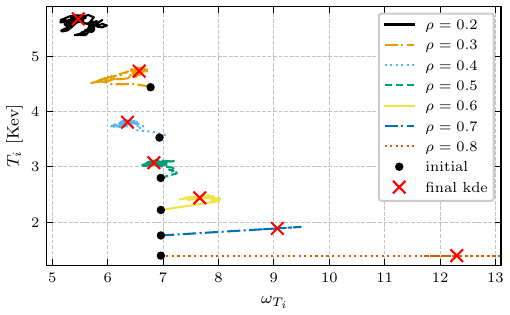}
    \caption{Evolution of the ion temperature $T_i$ and its normalized gradient $\omega_{T_i}$ for each of the seven radial positions during the adiabatic optimization. Initial conditions are marked by black dots; the final values selected via KDE are marked by red crosses.}
    \label{figure:updated_2}
\end{figure}

\begin{figure}
    \centering
    \includegraphics[width=0.90\linewidth]{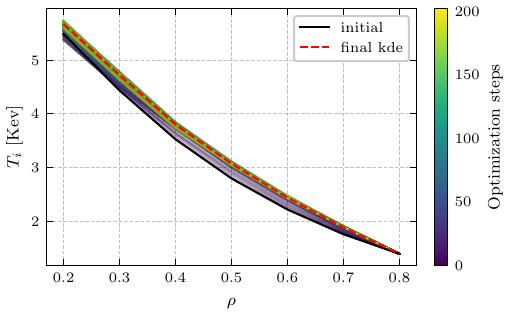}
    \caption{Ion temperature profiles during the adiabatic electrons optimization. Initial profile (black solid) and final KDE-estimated profile (red dashed) are highlighted.}
    \label{figure:temp_2_profile}
\end{figure}

\begin{figure}
    \centering
    \includegraphics[width=0.90\linewidth]{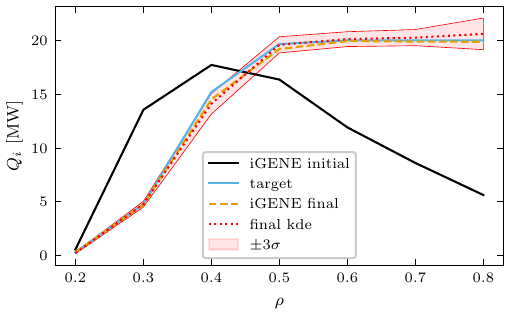}
    \caption{Ion heat flux profiles for the adiabatic electrons case. Black: initial profile. Light blue: target (integral of the heat source). Orange dashed: time-averaged flux during the converged optimization. Red dotted: post-optimization validation run at the final KDE-estimated profile.}
    \label{figure:flux_flow_2_profile}
\end{figure}

Figures~\ref{figure:flux_flow_2_trace} and~\ref{figure:updated_2} show that all seven radial points converge smoothly to their respective flux targets, and that the parameter trajectories cluster around stable equilibria despite turbulent noise.
The final parameters are extracted via the weighted KDE procedure described in Appendix~\ref{app:kde}, and the resulting temperature profile and heat flux profile (Figs.~\ref{figure:temp_2_profile}--\ref{figure:flux_flow_2_profile}) agree closely with the targets.
A post-optimization validation run confirms the result independently.

\begin{figure}
    \centering
    \includegraphics[width=0.90\linewidth]{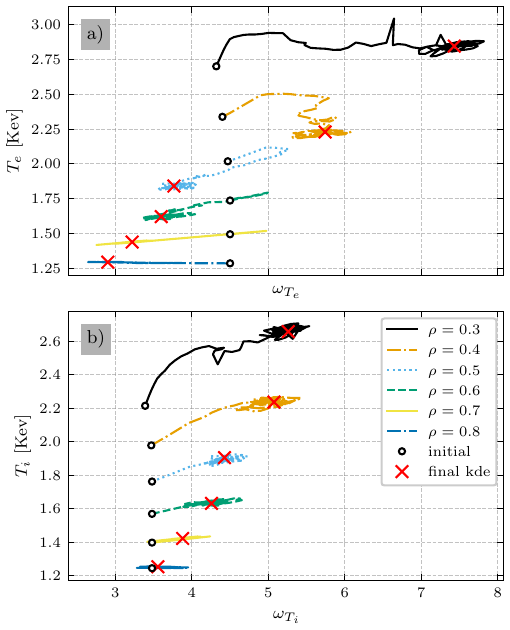}
    \caption{Parameter-space trajectories for the kinetic-electron case. (a) Electron temperature and gradient; (b) ion temperature and gradient. Notation as in Fig.~\ref{figure:updated_2}.}
    \label{figure:updated_E}
\end{figure}

\begin{figure}
    \centering
    \includegraphics[width=0.90\linewidth]{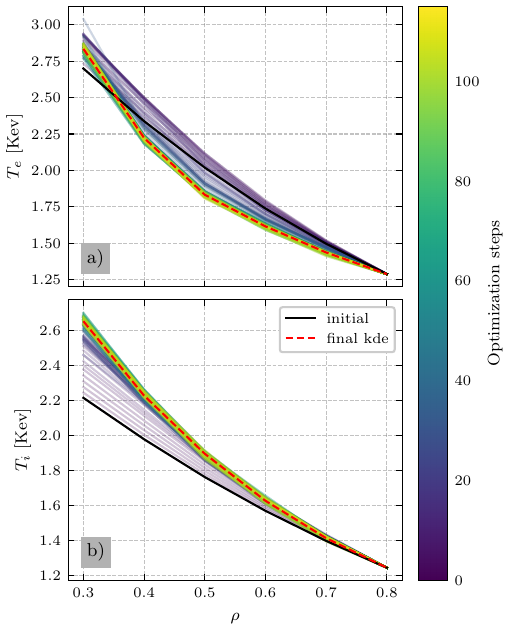}
    \caption{(a) Electron and (b) ion temperature profiles for the kinetic-electron optimization. Notation as in Fig.~\ref{figure:temp_2_profile}.}
    \label{figure:temp_E_profile}
\end{figure}

\begin{figure}
    \centering
    \includegraphics[width=0.90\linewidth]{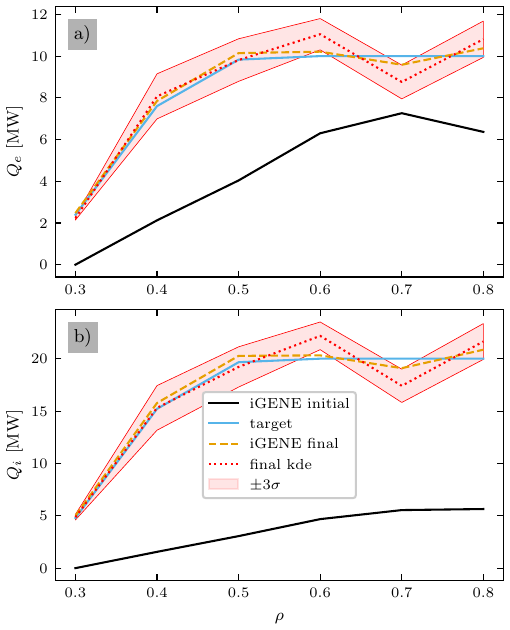}
    \caption{(a) Electron and (b) ion heat flux profiles for the kinetic-electron optimization. Notation as in Fig.~\ref{figure:flux_flow_2_profile}.}
    \label{figure:flow_E_profile}
\end{figure}

As a final test, we repeat the same optimization including also kinetic electrons. For simplicity, we assume heavy electrons, $m_e/m_i$ = 0.01 and evolve the temperature profiles only. We keep the same heat source as in the previous case  for the ions, while an additional 10 MW are injected in the electron channel. Simulations are performed in the electrostatic limit ($\beta_e=10^{-5}$). A mesh with ($n_x\times n_{k_y}\times n_z\times n_{v_\|} \times n_\mu$= $384\times48\times24\times64\times16$) is used; $L_x$ varies between $150\rho_s$ and $400\,\rho_s$ depending on the radial position (larger towards the edge). The simulations are run for a total time of $t=600\,t\frac{c_s}{a}$ and the gradients are computed with one backpropagation step. We keep the same boundary condition but shift the innermost point to $r/a = 0.3$ because at 0.2 the flux is essentially zero. This configuration requires 24 nodes with 8 GPUs (4 nodes) per radial simulation to satisfy the memory constraints. 
The optimization must now satisfy two coupled objectives simultaneously across six radial locations. The initial temperature profiles are rescaled compared to the used for adiabatic electrons in order to avoid nonphysically too large fluxes.
Figures~\ref{figure:updated_E}--\ref{figure:flow_E_profile} show that the optimizer is indeed able to satisfy both channels simultaneously and converges stably, with again the post-optimization validation run confirming the final profiles. We remark that the final profiles obtained in this case are significantly different from the one obtained with adiabatic electron and, somewhat unexpectedly, shows larger temperature gradients in the code compared to the edge. While this is not the typical situation in a real scenario, it is a consequence of both the simplified physical model we are considering, namely heavy electron and electrostatic collisionless limit, and the fixed density profile. The particle flux obtained for the final profiles is inward in the core, whereas we would expect a zero or positive one in an actual experiment. More details on the numerical convergence of the results are discussed in Appendix \ref{app:kin_el}.
The fact that cross-channel coupling does not destabilize the descent, despite the imperfect gradients and the turbulent noise in both channels, is encouraging for future applications to multi-species multi-channel transport. 


\section{Conclusion}
\label{sec:conclusion}

We have presented iGENE, a TensorFlow implementation of the local electromagnetic gyrokinetic model that enables automatic differentiation (AD) of linear and non-linear turbulent simulations. The code has been successfully validated against the standard version of GENE in several linear and nonlinear regimes. Gradients of quantities of interest can be be evaluated for both linear and nonlinear simulations. The former are accurate and reproduce traditional finite difference estimates. The latter can be computed as well but, due to their stochastic nature, are prone to numerical instabilities if they are evaluated over long backpropagation windows. Eventually  the Lyapunov instability develops and causes them to diverge. 

Nevertheless, we find that non-linear AD gradients are still  meaningful when evaluated within a finite backpropagation window whose extent correlates with the signal  autocorrelation time. As an example, for the case we have considered here, gradients of the heat fluxes with respect to $\omega_T$, $\omega_n$, and $q$ reach $15$--$34\%$ of their finite-difference values, and for $\epsilon$ approximately $50\%$. Despite these imperfections, these gradient are still useful and can be utilized for downstream optimization tasks. All the test we performed, namely single-point flux matching, seven-point adiabatic-electrons optimization, and six-point kinetic-electron optimization tasks, converge stably. Optimized results are confirmed by independent post-optimization validation runs. 

These results have several implications. First, we are able to compute on the fly gradients of many quantities of interest from a single high-fidelity gyrokinetic simulation. They can then be easily and directly used for downstream tasks such as uncertainty quantification or sensitivity analysis. For applications where only the gradient direction matters, such as the flux-matching optimization demonstrated here, the inaccurate non-linear AD gradients are sufficient, even when their magnitude deviates significantly from the finite-difference reference.
Finally, the divergence of the adjoint state for large $N$ is a known property of chaotic systems. To overcome this issue we have identified a practical rule for non-linear gyrokinetic turbulence by taking a window of order of the heat-flux autocorrelation time, typically a few hundred time steps. 
Whether strategies such as ensemble adjoint methods could extend this window further remains an open question.

Beyond the specific applications shown here, differentiable non-linear gyrokinetics opens several avenues: embedding the solver as a physics layer inside neural-network surrogate models~\cite{artigues2025accelerating}, Bayesian inference of profile parameters from experimental measurements, and sensitivity analysis for uncertainty quantification.
As the fusion community moves toward the design of burning-plasma devices, tools that combine first-principles fidelity with gradient-based accessibility will play an increasingly important role.

\appendix
\section*{Acknowledgments}
Computations were performed on the HPC system Viper-GPU at the Max Planck Computing and Data Facility.

\section{Parameter selection via weighted KDE}\label{app:kde}
\begin{figure}
    \centering
    \includegraphics[width=0.90\linewidth]{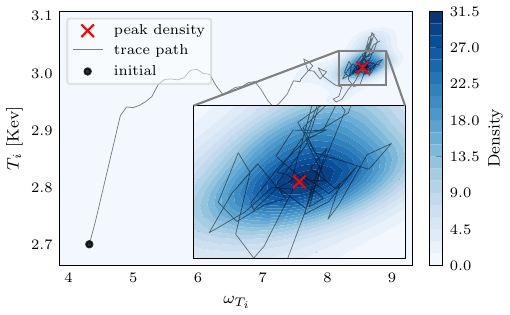}
    \caption{Illustration of the weighted KDE parameter selection procedure for a single radial position. The trajectory of the optimizer in the $(\omega_{T_i}, T_i)$ plane is shown. The background colormap shows the resulting KDE density estimate. The red cross marks the density maximum, which is taken as the final parameter value used to construct the temperature profile.}
    \label{figure:kde}
\end{figure}

Because gyrokinetic turbulence is a chaotic system, optimized parameters fluctuate about a statistical mean rather than settling to a fixed point. To extract representative final values we apply the following procedure. 
The optimization trajectory $\{(\omega_{T_i}^{(k)}, T_i^{(k)})\}$ is weighted linearly so that later iterates contribute more to the density estimate. 
A kernel density estimator (KDE) is then applied to the weighted dataset to generate a two-dimensional density map in parameter space. 
The final parameter values are taken as the location of the density maximum, which filters out high-frequency turbulent jitter and provides a stable, reproducible estimate of the converged equilibrium. 
An illustration of this procedure is given in Fig.~\ref{figure:kde}. 

\section{Convergence of kinetic electron simulation results}
\label{app:kin_el}
\begin{figure}
    \centering
    \includegraphics[width=0.90\linewidth]{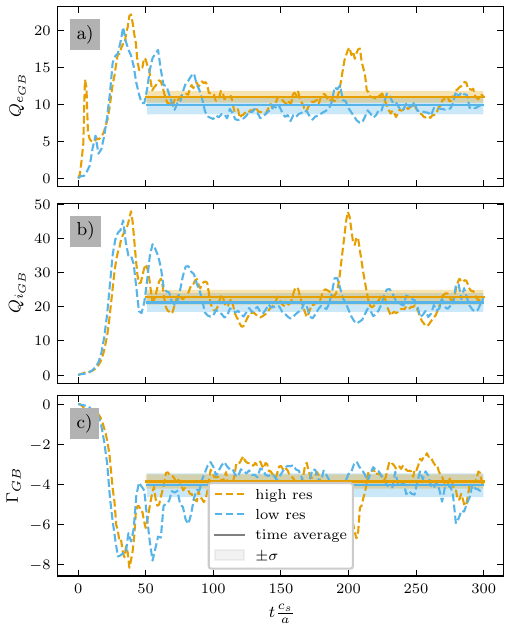}
    \caption{Comparison of time traces of heat and particle fluxes obtained from simulations using the same resolution as the one adopted for the profile optimization (labeled as low res.) and a higher one.}
    \label{figure:trace_converge}
\end{figure}

\begin{figure}
    \centering
    \includegraphics[width=0.90\linewidth]{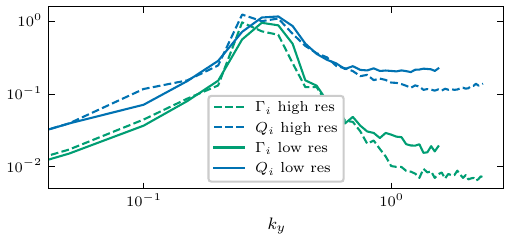}
    \caption{Comparison of flux spectra obtained with the resolution used for the profile optimization (labeled as low res.) and a higher resolution in all directions.}
    \label{figure:converge_spectra}
\end{figure}

We compare the simulation results obtained with the resolution used for the profile optimization with kinetic electrons ($n_x\times n_{k_y}\times n_z\times n_{v_\|} \times n_\mu $=$216\times32\times16\times48\times12$) to results obtained with a finer mesh ($n_x\times n_{k_y}\times n_z\times n_{v_\|} \times n_\mu$= $384\times48\times24\times64\times16$). Given the increased binormal resolution, a lower dissipation ($D_y$=0.01 instead of 0.025) was also used in the high resolution case). Figure\ref{figure:trace_converge} compares the time traces of the fluxes, showing that the resolution we used is sufficient to converge the transport through all channels. As can be seen in Figure \ref{figure:converge_spectra} the mesh is not completely resolved in the high $k_y$ region (where ETG modes are active) but the differences are minimal and justify the resolution adopted for the optimization.

\section*{Author declaration}

\subsection*{Conflict of interest}

The authors have no conflicts to disclose.

\section*{Data and Code Availability Statement}

The data that supports the findings of this study are available from the corresponding author upon reasonable request.

\section*{References}
\bibliographystyle{iopart-num}
\bibliography{bibliography}

\end{document}